\documentclass[final,5p,times,twocolumn]{elsarticle}

\usepackage{amssymb}
\usepackage{amsthm}
\usepackage{amsmath}
\usepackage{multirow}
\usepackage{textcomp}
\usepackage{subcaption} 
\usepackage{siunitx}
\usepackage{epstopdf}
\usepackage[table,xcdraw]{xcolor}
\usepackage{stmaryrd}
\usepackage{braket}    
\usepackage{siunitx}   

\journal{Journal of Optics and Laser Technology}

\begin{document}

\begin{frontmatter}



\title{ Optical Chopping Enhanced Rydberg-Atom-Based Ultra-Low-Frequency Electric Field Measurement}


\author[csu,ucas]{Yipeng Xie}
\ead{xieyipeng24@csu.ac.cn}
\affiliation[csu]{organization={Key Laboratory of Space Utilization, Technology and Engineering Center for Space Utilization, Chinese Academy of Sciences},
            postcode={100094},
            city={Beijing},
            country={China}}
\affiliation[ucas]{organization={University of Chinese Academy of Sciences},
            postcode={100049},
            city={Beijing},
            country={China}}

\author[csu]{Mingwei Lei}
\ead{mwlei@csu.ac.cn}
            
\author[csu]{Wenbo Dong}
\ead{wbdong@csu.ac.cn}       
\author[csu]{Meng Shi\corref{cor1}}
\ead{shimeng@csu.ac.cn}
\cortext[cor1]{Corresponding author}

\begin{abstract}
This study demonstrates a significant enhancement in ultra-low-frequency (ULF) electric field sensitivity using Rydberg atoms via an optical chopping amplification (OCA) technique. Conventional Rydberg-based ULF measurements are fundamentally limited by 1/f noise, which severely degrades sensitivity. Our approach modulates the coupling laser with an optical chopper before the vapor cell, inducing periodic Rydberg excitation at the chopping frequency. The photodetector (PD) output signal is demodulated by a lock-in amplifier (LIA) using the optical chopper's signal as the reference. This process effectively improves the signal-to-noise ratio (SNR) by shifting the 1/f noise to a higher frequency band where it can be filtered out. The OCA technique enhanced sensitivity by 19.1 dB for the frequency 7 Hz, which is down to 49.1 $\mu$V/cm/$\sqrt{\mathrm{Hz}}$. For the frequency range from 10Hz to1kHz, it also enhanced nearly 7dB. This OCA method for enhancing the sensitivity of Rydberg atoms in ULF electric field measurements enables the Rydberg sensor's detection range to span the entire spectrum from low frequency (LF) to ULF, thereby significantly broadening its application potential. 
\end{abstract}




\end{frontmatter}


\section{Introduction}
The measurement of ultra-low frequency (ULF) electric fields is critically important in several fields, including space physics\cite{Streltsov2018,Chen2018}, geophysics\cite{Njoku2003,Atkinson2006}, underwater information perception\cite{Li2025,Yang2025} and communication\cite{Kemp2019,Li2016}, owing to their long wavelengths and extensive propagation distances. However, the detection sensitivity of ULF receivers is typically constrained by 1/f noise, the primary source of which is active semiconductor devices, particularly the metal-oxide-semiconductor field-effect transistors (MOSFETs) employed for signal amplification and frequency mixing\cite{Sacchi2003}. In electronics, various techniques have been employed to mitigate 1/f noise, including lock-in amplification\cite{SCOFIELD1994}, complementary cascade switching\cite{Sanduleanu2025}, autozeroing\cite{Enz1996} and chopper stabilization\cite{ENZ1987}. 

Rydberg atoms have emerged as a promising technology for electric field sensing, offering several advantages over traditional sensors, such as high sensitivity, wide bandwidth, a large dynamic range, and a compact form factor\cite{Jing2020,Hu2022,Meyer2020}. Conventional ULF antennas are characterized by substantial physical dimensions. In contrast, Rydberg-atom sensors are not constrained by the traditional Chu limit\cite{Chu1948}, enabling miniaturization to the centimeter scale. However, the shielding effect of the atomic vapor cell on low-frequency (LF) electric field waves presents a significant challenge\cite{Jau2020,Abel2011} . This effect prevents the confined Rydberg atoms from detecting electromagnetic field components with frequencies below megahertz. Consequently, most current research on Rydberg-based electric field measurements focuses on the microwave band\cite{Bohaichuk2022,Anderson2021,Gordon2019,Yan2025}. Several research groups have employed Rydberg atoms for LF \cite{Jau2020,Li2023,Lei2024,Zhang2026} and shortwave\cite{Mao2024} electric field measurement. However, a critical limitation persists across these studies: none have successfully mitigated the 1/f noise inherent to the LF band, which fundamentally degrades measurement sensitivity.

In this work, we demonstrate a significant enhancement in the sensitivity of ULF electric field detection using Rydberg atoms via an optical chopping amplification (OCA) technique. The coupling laser is optically chopped before it enters the vapor cell. Consequently, the cesium atoms undergo periodic transitions between the first excited state and Rydberg state at the chopping frequency. To benchmark the performance of this approach, we compared the sensitivity of Rydberg sensors in ULF electric field measurements with and without using the OCA method. The results highlight the strengths of OCA method which can suppress the 1/f noise effectively.
\begin{figure}[t]
    \centering
    \includegraphics[scale=0.9]{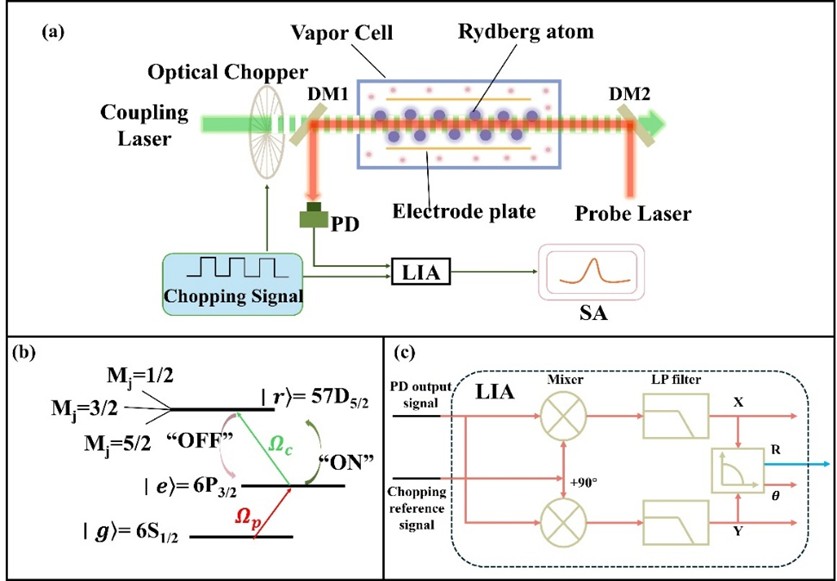}
    \caption{(a) Schematic of the experimental setup. An 852~nm probe laser propagates through a vapor cell. A pair of copper electrode plates is integrated into the cell in parallel with a spacing of 18~mm. A chopped 509~nm coupling laser counter-propagates and overlaps with the probe beam. The reference signal from the chopper and the output signal from the PD are fed into the LIA for demodulation. The resulting signal is measured by the SA. Abbreviations: DM, dichroic mirror; PD, photodetector; LIA, lock-in amplifier; SA, spectrum analyzer. (b) Energy-level diagram for the three-level Rydberg-EIT system. A probe laser resonantly couples the $\ket{6S_{1/2}}$ and $\ket{6P_{3/2}}$ states with Rabi frequency $\Omega_p$. A coupling laser drives the transition of $\ket{6P_{3/2}}$ and $\ket{57D_{5/2}}$ states with Rabi frequency $\Omega_c$. In the presence of an electric field, the Rydberg level exhibits $m_j = 1/2, 3/2, 5/2$ dependent Stark shifts and splitting. (c) The signal demodulation process within the LIA.}
    \label{fig:setup}
\end{figure}

\begin{figure}[t]
    \centering
    \includegraphics[scale=0.32]{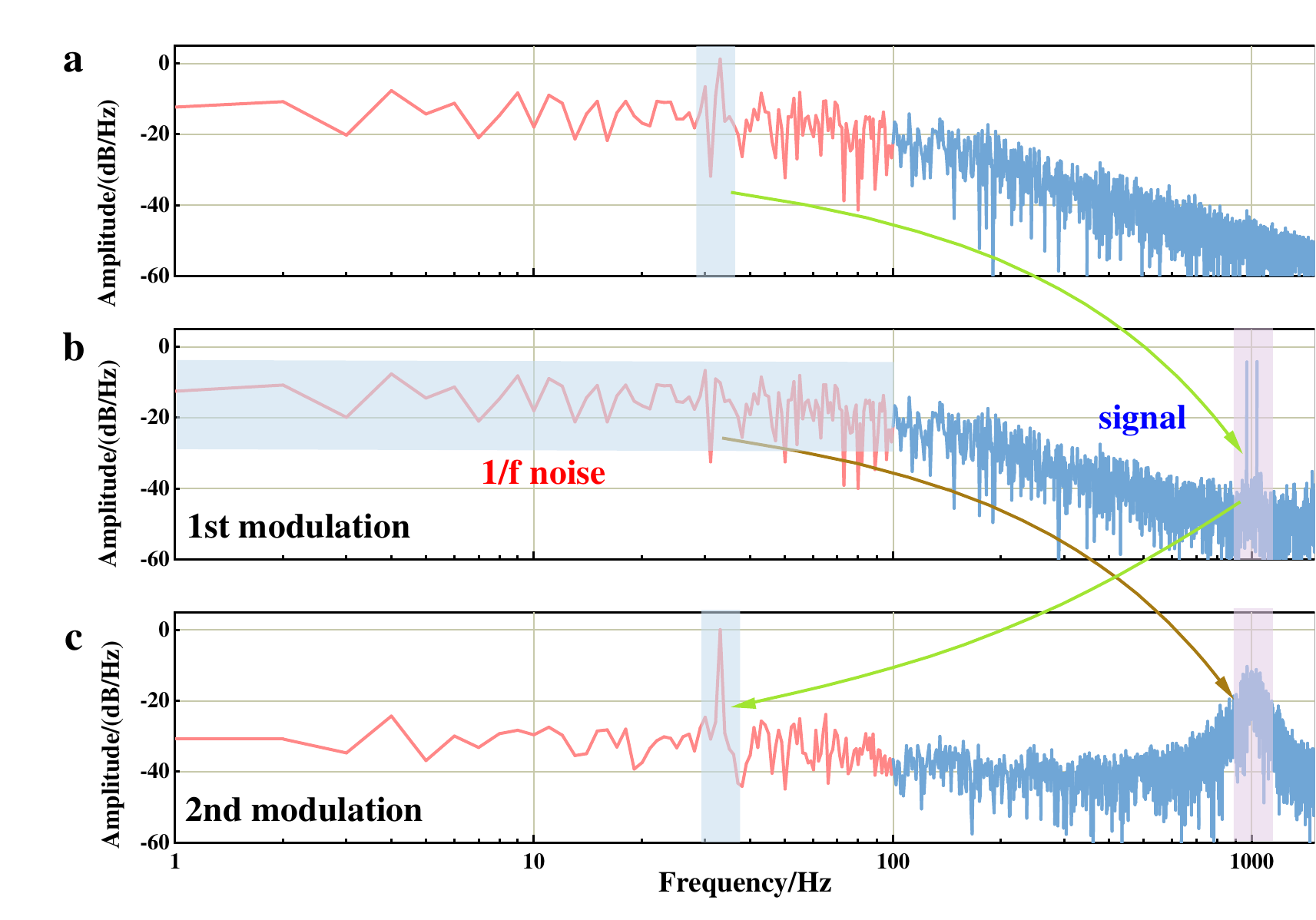}
    \caption{(a) Power spectral density without optical chopping, showing the ULF signal obscured by LF noise. (b) Spectrum after optical chopping at $f_{\mathrm{chop}}$. The signal is upconverted to the sidebands at $f_{\mathrm{chop}} \pm f_s$. (c) Spectrum after LIA demodulation. The signal is downconverted back to the baseband, while the $1/f$ noise introduced by the PD and transmission line is upconverted to the chopping frequency region and filtered out. The overall process involves double modulation of the signal and single modulation of the noise.}
    \label{fig:spectrum}  
\end{figure}

\label{sec:introduction}
\section{Method}

\subsection{Experimental Setup}
The experimental setup and corresponding three-level Rydberg-EIT diagram are illustrated in Fig.~1(a) and (b), respectively. The electric field is applied to the vapor cell with a diameter of 40~mm and a length of 100~mm using a pair of copper electrode plates with dimensions of $80~\mathrm{mm} \times 26~\mathrm{mm} \times 1~\mathrm{mm}$. The distance between the two electrodes is 18~mm. An 852~nm probe laser (TOPTICA) drives a transition of Cs atoms from the ground state $|6S_{1/2}\rangle$ to the excited state $|6P_{3/2}\rangle$ with a power of 500~$\mu$W. Simultaneously, a 509~nm coupling laser (TOPTICA) drives the transition of Cs atoms from $|6P_{3/2}\rangle$ to the Rydberg state $|57D_{5/2}\rangle$ with a power of 20~mW. Their $1/e^2$ beam waist $\omega_p$ and $\omega_c$ is 650 and 890~$\mu$m. Before entering the vapor cell, the coupling laser is modulated by an optical chopper at a frequency $f_{\mathrm{chop}}$. The reference signal from the chopper is input to the LIA along with the PD output signal for demodulation. The demodulated signal is measured using a spectrum analyzer (SA). In the presence of an electric field, the $|57D_{5/2}\rangle$ Rydberg level exhibits Stark shifts and splitting dependent on the $m_j$ quantum number ($m_j = 1/2, 3/2, 5/2$). The $m_j = 1/2$ state was selected for detection due to its higher polarizability compared to the $m_j = 3/2, 5/2$ states~\cite{Li2023}. To suppress laser frequency noise, the frequencies of both the probe and coupling lasers are stabilized by locking them to an ultra-stable optical cavity using the Pound-Drever-Hall (PDH) technique. The optical chopper modulates the coupling laser, thereby controlling the system's quantum state. During the ``OFF'' phase of the coupling laser, the atomic system decays from the Rydberg state $|57D_{5/2}\rangle$ to the intermediate excited state $|6P_{3/2}\rangle$. During the ``ON'' phase of the coupling laser, the system is re-excited from $|6P_{3/2}\rangle$ back to the Rydberg state $|57D_{5/2}\rangle$.

Fig.~1(c) schematically illustrates the signal demodulation process within the LIA. Inside the LIA, the output signal from the PD is split into two paths, each fed into a separate mixer. One signal path is mixed directly with the chopping reference signal. The other path is first phase-shifted by 90$^\circ$ and then mixed with the same reference signal. The outputs from both mixers are passed through individual low-pass filters (LPFs) to attenuate high-frequency components, yielding the in-phase ($X$) and quadrature ($Y$) signals. These $X$ and $Y$ signals are then processed by a vector operation module, which calculates the resultant amplitude ($R$) and phase ($\theta$).

\subsection{Optical Chopping Amplification}

When measuring ULF signals using the AC Stark effect, the Rydberg state $\ket{57D_{5/2}}$ undergoes a shift under the influence of the signal field and its Rabi frequency shows a corresponding relationship with the intensity of the signal field:
\begin{equation}
\Delta f_{\mathrm{Stark}} = -\frac{1}{2}\alpha E_{\mathrm{tot}}^2
\label{eq:stark1}
\end{equation}
herein, $\alpha$ represents the polarizability of atomic energy levels with respect to the signal electric field, and $E_{\mathrm{tot}}$ denotes the total electric field applied to the atoms. In the scenario of receiving time-harmonic signals, where $E_{\mathrm{tot}} = A_{\mathrm{sig}}\cos(\omega_{\mathrm{sig}}t + \varphi_{\mathrm{sig}})$. The distance of the energy level shift exhibits a quadratic relationship with the amplitude of the electric field. When an auxiliary DC field is introduced, $E_{\mathrm{tot}}^2$ transforms into ($E_{\mathrm{DC}} \gg E_{\mathrm{sig}}$):
\begin{equation}
E_{\mathrm{tot}}^2 \approx E_{\mathrm{DC}}^2 + 2E_{\mathrm{DC}}E_{\mathrm{sig}}\cos(\omega_{\mathrm{sig}}t + \varphi_{\mathrm{sig}})
\label{eq:etot}
\end{equation}
the corresponding Rabi frequency of the Stark shift is:
\begin{equation}
\Delta f_{\mathrm{Stark}} = -\frac{\alpha E_{\mathrm{DC}}^2}{2} - \alpha E_{\mathrm{DC}}E_{\mathrm{sig}}\cos(\omega_{\mathrm{sig}}t + \varphi_{\mathrm{sig}})
\label{eq:stark2}
\end{equation}
When measuring ULF signals, the auxiliary DC field can act as an enhancement factor to boost the Rabi frequency of energy level shifts and convert the second-order response into a first-order linear response. After locking the coupling laser frequency at a designated position on the spectrum, the Stark shift generated by atomic energy levels can be converted into a change in optical power, thereby producing a voltage signal at the output of the photodetector:
\begin{equation}
V_{\mathrm{PD}}(t) = V_s(t) + V_n(t) = \beta\Delta f_{\mathrm{Stark}} + V_n(t)
\label{eq:vpd}
\end{equation}
here, $\beta$ represents the slope of the EIT spectrum at the locked frequency point of the coupling laser, $V_s(t)$ denotes the signal, $V_n(t)$ represents the $1/f$ noise caused by active semiconductor devices.

In this study, the OCA technique used for ULF electric field measurement with Rydberg atoms operates on a principle analogous to electronic chopping amplification. Specifically, it translocates the $1/f$ noise and the signal of interest into distinct frequency bands. The entire process of optical chopping-enhanced electric field measurement is illustrated intuitively by spectrum shown in Fig.~2. An electric field signal and the optimal DC bias (the frequency of signal is $f_s$) were applied to the internal electrode plates of the vapor cell without using OCA technique, the PD output spectrum is shown in Fig.~2(a), which is severely degraded by $1/f$ noise. The output spectrum of the PD $S_{\mathrm{PD}}(f)$ (Fourier transform of $V_{\mathrm{PD}}(t)$) is shown as follow, where $S_s(f)$ denotes the signal spectrum, $S_n(f)$ represents the $1/f$ noise spectrum, and $k$ is a constant.
\begin{equation}
S_{\mathrm{PD}}(f) = S_s(f) + S_n(f), \quad \left(S_n(f) = \frac{k}{|f|}\right)
\label{eq:spd}
\end{equation}
When the optical chopper is in operation, it is driven by a periodic square wave at a frequency $f_{\mathrm{chop}}$. To simplify the analysis, only the fundamental component of its Fourier-series expansion $M(f)$ is considered (Eq.~\eqref{eq:mf}) and any gain factors arising in the overall process are neglected. Consequently, when the chopper modulates the coupling laser, the signal spectrum is convolved with the spectral function of this fundamental component, whereas the noise introduced by the PD remains unmodulated (Eq.~\eqref{eq:spd2}). Therefore, the signal is upconverted to the sidebands at $f_{\mathrm{chop}} \pm f_s$. The resulting optically chopped signal spectrum is shown in Fig.~2(b).
\begin{equation}
\begin{aligned}
M(f) = \mathcal{F}\{\cos(2\pi f_{\mathrm{chop}}t)\} 
= \frac{1}{2}\Big[\delta(f - f_{\mathrm{chop}}) + \delta(f + f_{\mathrm{chop}})\Big]
\end{aligned}
\label{eq:mf}
\end{equation}

\begin{equation}
\begin{aligned}
S_{\mathrm{PD}}'(f) &= S_s(f) * M(f) + S_n(f) \\
&= \frac{1}{2}\Big[S_s(f - f_{\mathrm{chop}}) + S_s(f + f_{\mathrm{chop}})\Big] + S_n(f)
\end{aligned}
\label{eq:spd2}
\end{equation}
The PD output and the chopper reference signal are fed into the LIA for synchronous demodulation. Consequently, shown as Fig.~2(c), the signal is recovered in the baseband, whereas the $1/f$ noise is upconverted to the chopping frequency region (Eq.~\eqref{eq:sde}) and suppressed by the LPF within the LIA.

\begin{equation}
\begin{aligned}
S_{\mathrm{de}}(f) &= S_{\mathrm{PD}}'(f) * M(f) \\
&= S_s(f) + \frac{1}{2}\left[S_n(f - f_{\mathrm{chop}}) + S_n(f + f_{\mathrm{chop}})\right]
\end{aligned}
\label{eq:sde}
\end{equation}

Through this process, the ULF electric field signal is modulated twice, effectively down-converting it to a baseband frequency. In contrast, the $1/f$ noise undergoes only a single modulation step, shifting it to a higher frequency band near the chopping frequency $f_{\mathrm{chop}}$. The signal is thus separated from the noise by the subsequent action of the low-pass filter (LPF), which passes the baseband signal while rejecting the higher-frequency noise.

\subsection{Atomic System}

A three-level atomic system of cesium atoms involving the ground state $\ket{g}$, the first excited state $\ket{e}$, and a Rydberg state $\ket{r}$. A probe laser, resonantly coupling states $\ket{g}$ and $\ket{e}$, drives the atomic transition with a constant Rabi frequency $\Omega_p$. Simultaneously, a coupling laser, resonantly connecting states $\ket{e}$ and $\ket{r}$, is periodically modulated by an optical chopper with frequency $f_{\mathrm{chop}}$. The coupling laser's Rabi frequency is therefore time-dependent, exhibiting a square-wave profile. Under the dipole and rotating-wave approximations, the system Hamiltonian in the interaction picture can be expressed as:
\begin{figure*}
    \centering
    \includegraphics[width=0.8\textwidth]{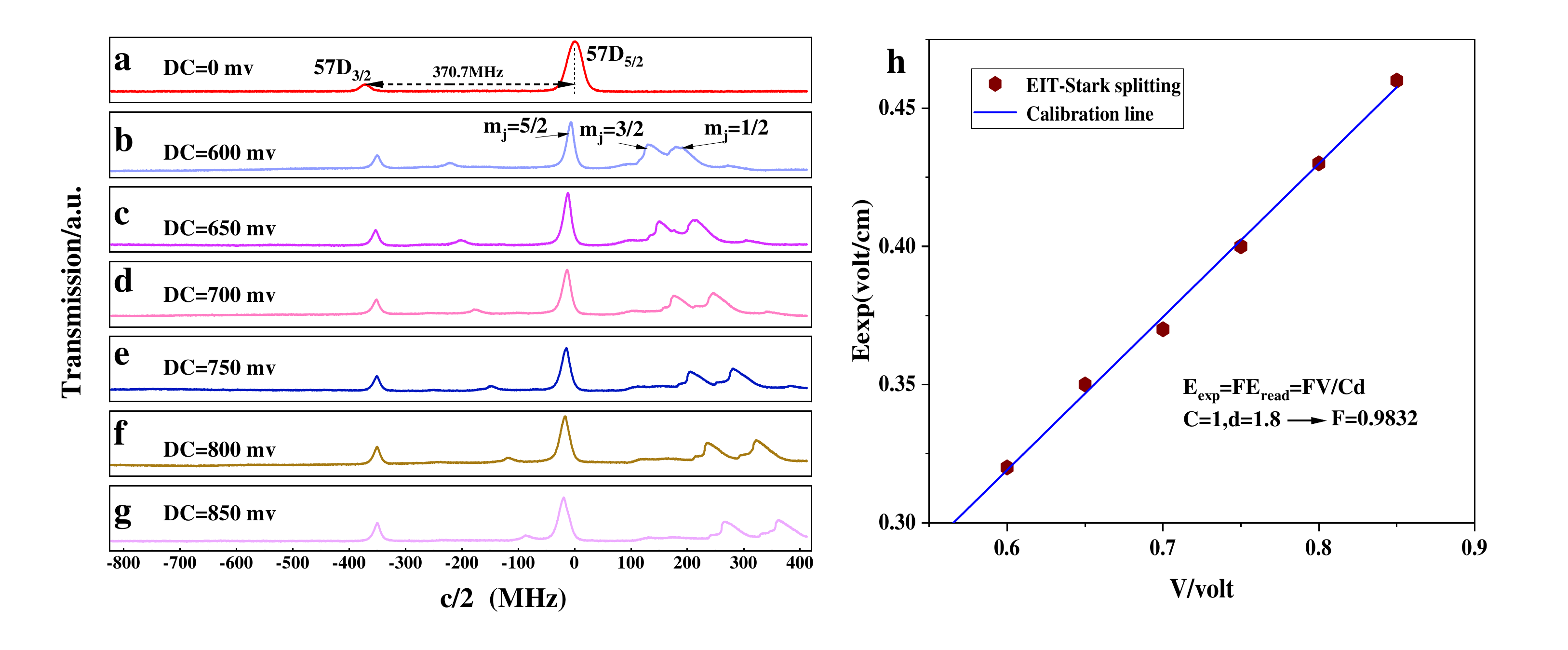}
    \caption{(a) to (g) represent the energy-level splitting of the EIT spectra when the direct DC electric field applied between the parallel electrodes is 0~mV, 600~mV, 650~mV, 700~mV, 750~mV, 800~mV, and 850~mV, respectively. (h) Calibration of the transmission factor when measuring the 66~Hz electric field in this work. The factor $F$ was determined by a linear fit of the electric field strength measured via the Rydberg atom response (using the $\ket{57D_{5/2}, m_j = 1/2}$ Stark shift) to the nominal field strength set by the signal generator, under strong-field conditions.}
    \label{fig:cali}  
\end{figure*}
\begin{equation}
H = \frac{\hbar\Omega_p}{2}\left(\ket{g}\bra{e} + \ket{e}\bra{g}\right) + \frac{\hbar\Omega_c(t)}{2}\left(\ket{e}\bra{r} + \ket{r}\bra{e}\right)
\label{eq:hamiltonian}
\end{equation}
here, the time-dependent coupling Rabi frequency $\Omega_c(t)$ is defined by:
\begin{equation}
\Omega_c(t) = 
\begin{cases}
\Omega_{c0} & \text{for } nT \leq t < nT + T/2 \\
0 & \text{for } nT + T/2 \leq t < (n+1)T
\end{cases}
\label{eq:omega_c}
\end{equation}
where $T = 1/f_{\mathrm{chop}}$ is the chopping period, and $\Omega_{c0}$ represents the maximum Rabi frequency of the coupling laser. This periodic modulation of the coupling laser forms the fundamental driving mechanism for the observed cyclic population transfer between the excited and Rydberg states. To accurately model the system's dynamics, we must account for spontaneous decay processes. The Rydberg state $\ket{r}$ decays to the first excited state $\ket{e}$ with a rate $\Gamma_r$, while the excited state $\ket{e}$ decays to the ground state $\ket{g}$ with a rate $\Gamma_e$. The system's evolution is therefore described by the density matrix $\rho$, which obeys the master equation:
\begin{equation}
\frac{d\rho}{dt} = -\frac{i}{\hbar}[H, \rho] + \mathcal{L}[\rho]
\label{eq:master}
\end{equation}
The Lindblad operator $\mathcal{L}[\rho]$ incorporates the spontaneous emission effects:
\begin{equation}
\mathcal{L}[\rho] = \frac{\Gamma_e}{2}\left(2\sigma_{ge}\rho\sigma_{eg} - \sigma_{ee}\rho - \rho\sigma_{ee}\right) + \frac{\Gamma_r}{2}\left(2\sigma_{er}\rho\sigma_{re} - \sigma_{rr}\rho - \rho\sigma_{rr}\right)
\label{eq:lindblad}
\end{equation}
where we have defined the lowering operators $\sigma_{ge} = \ket{g}\bra{e}$, $\sigma_{eg} = \ket{e}\bra{g}$, $\sigma_{er} = \ket{e}\bra{r}$, and $\sigma_{re} = \ket{r}\bra{e}$, with the projection operators $\sigma_{ii} = \ket{i}\bra{i}$ for $i \in \{e, r\}$. Expanding the master equation yields the optical Bloch equations for the density matrix elements, which form a complete set describing the population dynamics and coherence evolution under the combined influence of coherent laser drives and spontaneous decay processes. The periodic modulation of the coupling laser creates a distinct dynamical regime characterized by alternating intervals of coherent driving and spontaneous decay. During the ``ON'' phase of the coupling laser ($nT \leq t < nT + T/2$), when $\Omega_c(t) = \Omega_{c0}$, the system experiences coherent dynamics between states $\ket{e}$ and $\ket{r}$. For an atom initially in state $\ket{e}$, the coherent evolution in the absence of decay would follow:
\begin{equation}
\ket{\psi(t)} = \cos\left(\frac{\Omega_{c0}t}{2}\right)\ket{e} - i\sin\left(\frac{\Omega_{c0}t}{2}\right)\ket{r}
\label{eq:psi_t}
\end{equation}
with corresponding populations $P_e(t) = \cos^2(\Omega_{c0}t/2)$ and $P_r(t) = \sin^2(\Omega_{c0}t/2)$. However, in the presence of spontaneous emission, this coherent oscillation is damped, and the system approaches a steady state determined by the balance between driving and decay. During the ``OFF'' phase of the coupling laser ($nT + T/2 \leq t < (n+1)T$), when $\Omega_c(t) = 0$, the coherent coupling between $\ket{e}$ and $\ket{r}$ vanishes. The populations then evolve according to:

\begin{equation}
\rho_{rr}(t) = \rho_{rr}(t_0)e^{-\Gamma_r(t-t_0)}
\label{eq:rho_rr}
\end{equation}
\begin{equation}
\rho_{ee}(t) = \rho_{ee}(t_0) + \rho_{rr}(t_0)\left(1 - e^{-\Gamma_r(t-t_0)}\right) - \int_{t_0}^{t}\Gamma_e\rho_{ee}(\tau)\,d\tau
\label{eq:rho_ee}
\end{equation}
\begin{equation}
\rho_{er}(t) = \rho_{er}(t_0)e^{-(\Gamma_r+\Gamma_e)(t-t_0)/2}
\label{eq:rho_er}
\end{equation}
where $t_0 = nT + T/2$ marks the beginning of the ``OFF'' phase. The coherences decay exponentially, while the Rydberg population transfers to the excited state through spontaneous emission at rate $\Gamma_r$. The probe laser continuously pumps population from the ground state to the excited state, ensuring a reservoir of atoms in $\ket{e}$ available for excitation to the Rydberg state when the coupling laser is turned on. For the cyclic process to operate efficiently, the chopping period $T$ should be 
comparable to or longer than the spontaneous decay times, specifically $T \geq 1/\Gamma_r$, ensuring complete depletion of the Rydberg population during the ``OFF'' phase. Additionally, the relationship between the coupling Rabi frequency and the chopping duration determines the maximum Rydberg excitation achievable during each ``ON'' cycle.

\subsection{Calibration of Electric Field Measurement}

The curve in Fig.~3(a) illustrates the EIT spectrum in the absence of an external field, obtained by scanning the detuning of the coupling laser ($\Delta_c$); this spectrum serves as a reference. The detuning axis is calibrated relative to the fine structure components of the $\ket{57D_j}$ state~\cite{Jiao2017}. The curves in Fig.~3(b)--(g) show the evolution of EIT spectral splitting for applied DC electric fields of 0, 600, 650, 700, 750, 800, and 850~mV, respectively. The fine structure splitting between the $\ket{57D_{5/2}}$ and $\ket{57D_{3/2}}$ states in the absence of a field, as calculated using the ARC Python library, is 370.7~MHz (Fig.~3(a)). The application of the DC field lifts the degeneracy of the $m_j$ magnetic substates, leading to $m_j$-dependent Stark shifts and splittings for the states with $m_j = 1/2, 3/2$, and $5/2$. As reported in Reference~\cite{Li2023}, the $m_j = 1/2$ substate exhibits a larger polarizability and, consequently, greater sensitivity to the electric field.
\begin{figure*}
    \centering
    \includegraphics[width=0.8\textwidth]{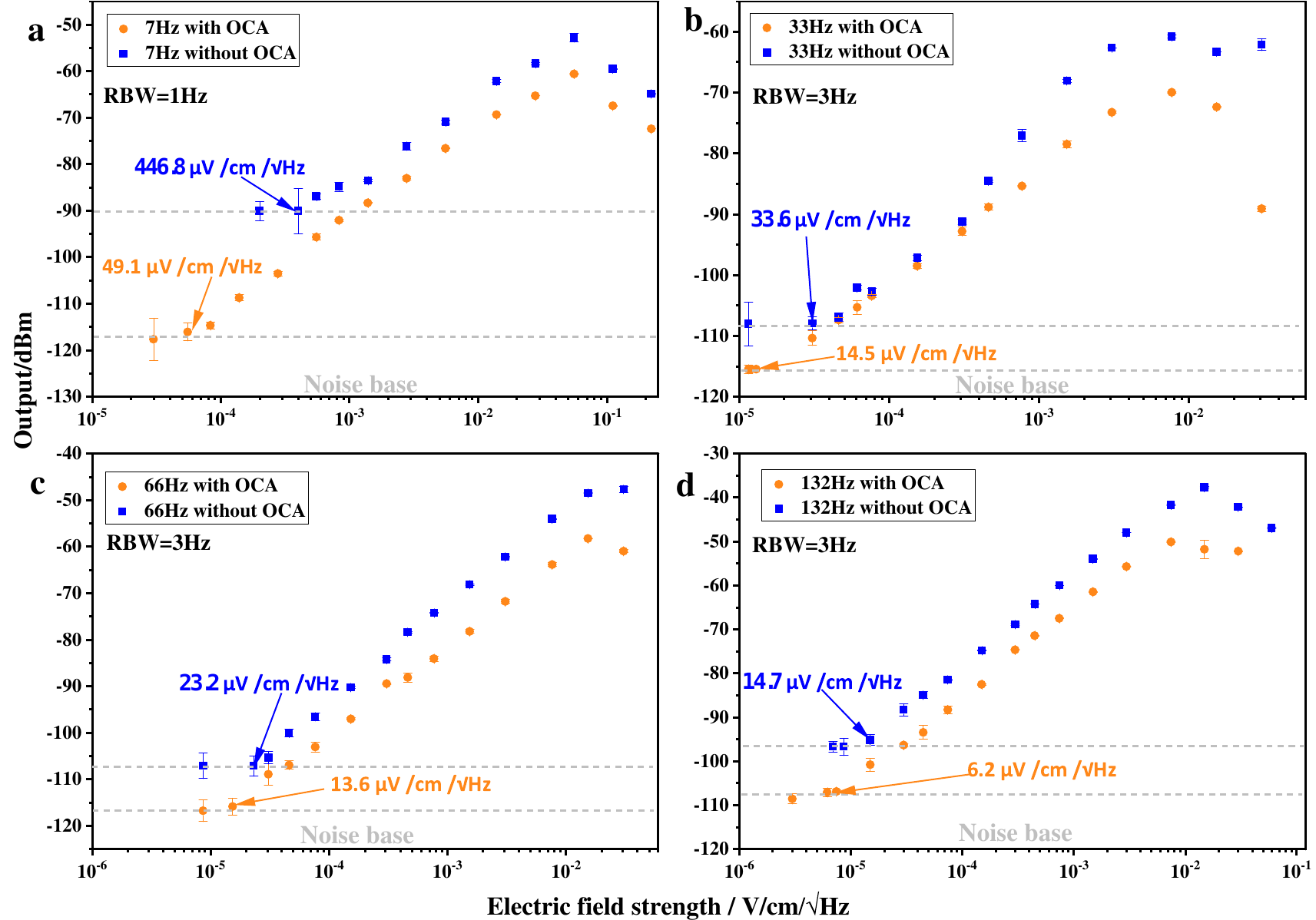}
    \caption{Comparison of ULF electric field measurements with and without using OCA. (a)--(d) Electric field strength measured at 7~Hz, 33~Hz, 66~Hz, and 132~Hz.}
    \label{fig:duibi}  
\end{figure*}
\begin{figure*}[t]
    \centering
    \includegraphics[width=0.8\textwidth]{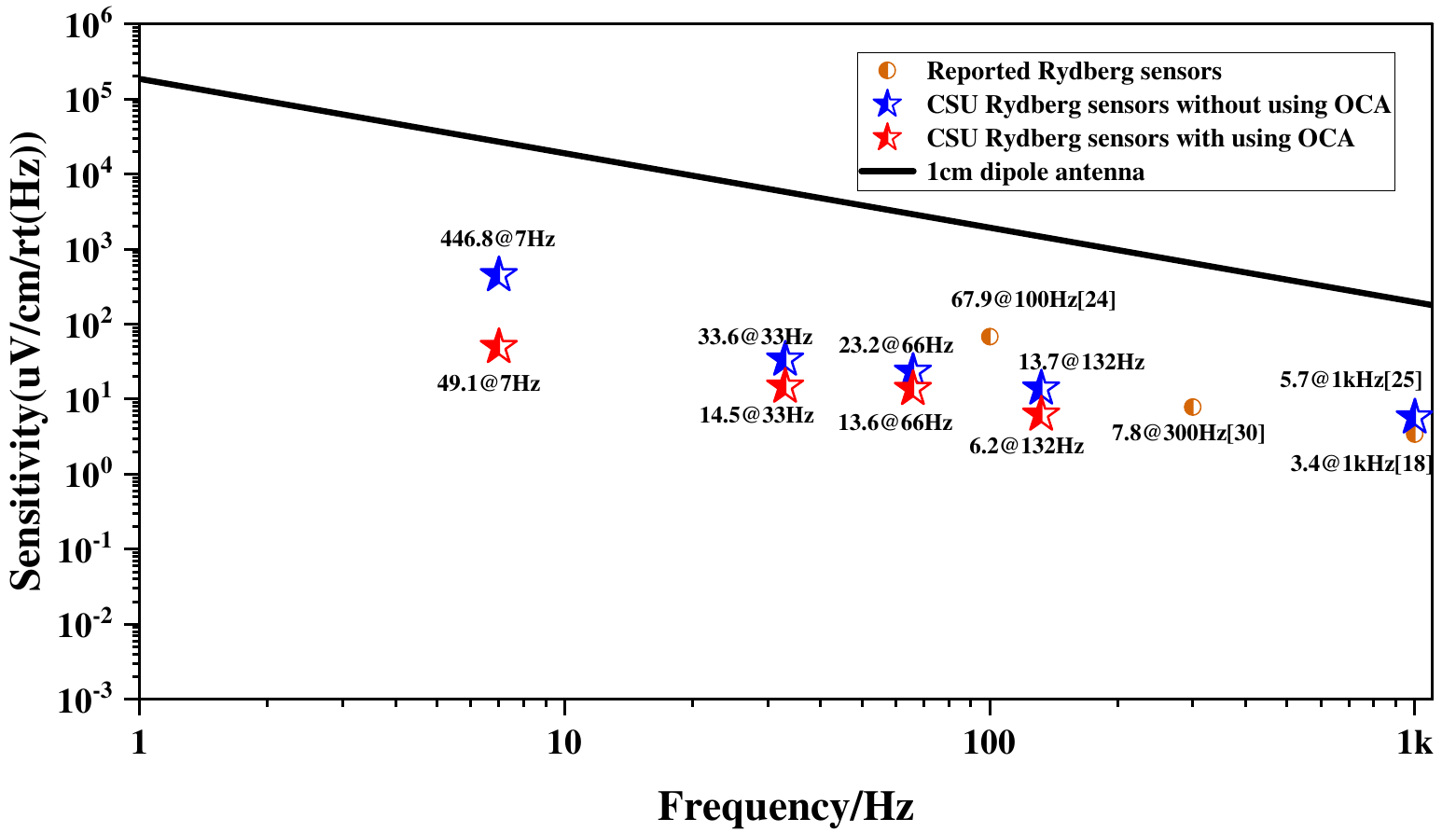}
    \caption{ Electric field sensitivity versus frequency. A comparison of the theoretical limit for a 1 cm dipole antenna with experimentally reported sensitivities from this work (using and without using OCA) and other Rydberg-atom-based experiments.}
    \label{fig:quanjuduibi}  
\end{figure*}
The strength of the signal field is defined as $E_{\mathrm{sig}} = FE_{\mathrm{read}}$. Here, $E_{\mathrm{read}}$ is calculated as $V/(Cd)$, where $V$ is the voltage (in volts) displayed on the signal generator, $d$ is the distance between the two electrodes, and $C$ is an effective voltage coefficient, with $C = \sqrt{2}$ for an AC signal field and $C = 1$ for a DC field. The calibration transmission factor, $F$, is determined from the equation $F = E_{\mathrm{exp}}/E_{\mathrm{read}}$, which is derived from measurements in a strong-field region. The experimental field strength, $E_{\mathrm{exp}}$, is obtained by measuring the Stark shifts of the $m_j = 1/2$ EIT signal, as shown in Fig.~3(b)--(g). The Stark shift is given by Eq.~\eqref{eq:stark1} and the polarizability $\alpha$ of the $\ket{57D_{5/2}, m_j = 1/2}$ state calculated using ARC package~\cite{Sibalic2017} is approximately $-3426.5~\mathrm{MHz~cm^2~V^{-2}}$. The calibration factor $F$ was subsequently determined from a linear fit of the locally measured field strengths to the applied voltages from the signal generator under strong-field conditions. Fig.~3(h) shows the calibration of the transmission factor when measuring the 66~Hz electric field in this work, based on the aforementioned calculation, $F$ for this measurement was determined to be 0.9832.

\label{sec:method}

\section{Results and discussion}

In the ULF band, discrete frequencies of 7~Hz, 33~Hz, 66~Hz, and 132~Hz were selected to avoid interference from the 50~Hz power-line fundamental and its harmonics, which typically occur at integer multiples of 5 or 10~Hz. These four frequencies were used to compare the electric field sensitivity achieved with and without the OCA technique. Fig.~4(a)--(d) present the electric field measurements at 7~Hz, 33~Hz, 66~Hz, and 132~Hz, respectively, with each subplot showing comparative results obtained with and without the OCA technique. For the measurements at 33~Hz, 66~Hz, and 132~Hz, the RBW of the spectrum analyzer was fixed at 3~Hz, and the DC bias fields applied to the Rydberg vapor cell were optimized to 590~mV, 580~mV, and 680~mV, respectively. The corresponding calibration factors $F$ were 0.9568, 0.9832, and 0.9304. The electric-field sensitivities achieved with the OCA technique at these three frequencies are 14.5, 13.6, and 6.2~$\mathrm{\mu V/cm/\sqrt{Hz}}$, representing enhancements of 7.3~dB, 4.6~dB, and 7.0~dB compared to the conventional approach.

The detection of weak electric fields at 7~Hz is particularly challenging due to the significant $1/f$ noise in this frequency region. To maximize sensitivity, the RBW was reduced to 1~Hz, and the DC bias was optimized to 590~mV with a calibration factor $F$ of 0.9975. As shown in Fig.~4(a), the OCA technique achieves a sensitivity of 49.1~$\mathrm{\mu V/cm/\sqrt{Hz}}$ at 7~Hz, compared to 446.8~$\mathrm{\mu V/cm/\sqrt{Hz}}$ without OCA. This represents a sensitivity enhancement of 19.1~dB, or approximately one order of magnitude. This substantial improvement is attributed to the OCA technique's ability to upconvert the dominant $1/f$ noise from the ULF band to the chopping frequency, thereby suppressing the noise floor at the signal frequency.

Fig.~5 compares the electric-field sensitivity achieved in this work with those reported in other Rydberg-atom-based experiments below 1~kHz~\cite{Jau2020,Li2023,Lei2024,Arumugam2025}. The solid curve represents the theoretical sensitivity limit of a 1~cm dipole antenna as a function of frequency~\cite{Hansen2006}. To the best of our knowledge, no prior Rydberg-based measurement has demonstrated sensitivity below 100~Hz, let alone 10~Hz. Our work thus bridges a critical gap in ULF electrometry, highlighting the exceptional performance of OCA approach for ULF electric-field detection with Rydberg atoms.

\label{sec:result}
\section{Conclusions}

In summary, we have demonstrated an enhancement in the sensitivity of ULF electric field detection by implementing an OCA technique with Rydberg atoms. The OCA technique suppresses the dominant $1/f$ noise by modulating the atomic response at a high-frequency carrier (via an optically chopped coupling laser) and subsequently recovering the signal through synchronous demodulation. This method enables the detection of weak ULF electric fields with sensitivities substantially surpassing those of conventional, direct-detection schemes. Utilizing the OCA method, a sensitivity enhancement of 19.1~dB was achieved at 7~Hz, which corresponds to a measured sensitivity of 49.1~$\mathrm{\mu V/cm/\sqrt{Hz}}$. Furthermore, an average improvement of approximately 7~dB was observed across the 10~Hz to 1~kHz range. These results establish a new benchmark for ULF electrometry and underscore the exceptional potential of Rydberg-atom-based sensors for high-sensitivity measurements.

The OCA technique paves the way for applications in fields such as geophysical surveying, sub-surface communication and underwater communication, where precise ULF electric field detection is paramount. Future work will focus on optimizing system integration, exploring advanced noise-reduction strategies, and extending the operational bandwidth to achieve even higher sensitivity across a broader frequency range. The significant sensitivity improvements demonstrated here, combined with the inherent potential for miniaturization of Rydberg sensors, indicate a clear path toward the development of compact, field-deployable devices for practical ULF electrometry.



\section*{Declaration of competing interest}
The authors declare that they have no known competing financial interests or personal relationships that could have appeared to influence the work reported in this paper.
\section*{Acknowledgements}

This work was supported by the National Key Research and Development Program of China (Grant No. 2024YFB3909500), the National Natural Science Foundation of China (Grant No. U2341211), the National Natural Science Foundation of China (Grant No. 62501568).
\section*{Data availability}
Data will be made available on request.

\bibliographystyle{elsarticle-num-names} 
\bibliography{cas-refs}

@article{ Streltsov2018,
Author = {Streltsov, A. V. and Berthelier, J. -J. and Chernyshov, A. A. and
   Frolov, V. L. and Honary, F. and Kosch, M. J. and McCoy, R. P. and
   Mishin, E. V. and Rietveld, M. T.},
Title = {Past, Present and Future of Active Radio Frequency Experiments in Space},
Journal = {SPACE SCIENCE REVIEWS},
Year = {2018},
Volume = {214},
Number = {8},
Month = {DEC},
DOI = {10.1007/s11214-018-0549-7},
Article-Number = {118},
ISSN = {0038-6308},
EISSN = {1572-9672},
ResearcherID-Numbers = {Frolov, Vladimir/AAU-9780-2020
   Chernyshov, Alexander/F-1748-2017},
ORCID-Numbers = {Honary, Farideh/0000-0002-7998-5628
   MISHIN, EVGENY/0000-0002-3183-0600
   Kosch, Michael Jurgen/0000-0003-2846-3915
   Chernyshov, Alexander/0000-0003-4692-9643},
Unique-ID = {WOS:000448857400001},
}

@article{ Chen2018,
Author = {Chen, Linjie and Aminaei, Amin and Gurvits, Leonid I. and Wolt, Marc
   Klein and Pourshaghaghi, Hamid Reza and Yan, Yihua and Falcke, Heino},
Title = {Antenna design and implementation for the future space Ultra-Long
   wavelength radio telescope},
Journal = {EXPERIMENTAL ASTRONOMY},
Year = {2018},
Volume = {45},
Number = {2},
Pages = {231-253},
Month = {APR},
DOI = {10.1007/s10686-018-9576-3},
ISSN = {0922-6435},
EISSN = {1572-9508},
ResearcherID-Numbers = {Yan, Yihua/AGY-9819-2022
   Aminaei, Amin/AAR-4344-2021
   },
ORCID-Numbers = {Yan, Yihua/0000-0002-7106-6029
   Chen, Linjie/0000-0002-3618-3430
   Gurvits, Leonid/0000-0002-0694-2459},
Unique-ID = {WOS:000430300100006},
}

@article{ Njoku2003,
Author = {Njoku, EG and Jackson, TJ and Lakshmi, V and Chan, TK and Nghiem, SV},
Title = {Soil moisture retrieval from AMSR-E},
Journal = {IEEE TRANSACTIONS ON GEOSCIENCE AND REMOTE SENSING},
Year = {2003},
Volume = {41},
Number = {2},
Pages = {215-229},
Month = {FEB},
DOI = {10.1109/TGRS.2002.808243},
ISSN = {0196-2892},
EISSN = {1558-0644},
ResearcherID-Numbers = {Lakshmi, Venkat/I-3078-2016},
Unique-ID = {WOS:000182494600006},
}

@article{ Atkinson2006,
Author = {Atkinson, Gail M. and Boore, David M.},
Title = {Earthquake ground-motion prediction equations for eastern North America},
Journal = {BULLETIN OF THE SEISMOLOGICAL SOCIETY OF AMERICA},
Year = {2006},
Volume = {96},
Number = {6},
Pages = {2181-2205},
Month = {DEC},
DOI = {10.1785/0120050245},
ISSN = {0037-1106},
EISSN = {1943-3573},
ResearcherID-Numbers = {Boore, David/GWQ-6027-2022},
Unique-ID = {WOS:000242821700016},
}

@article{ Li2025,
Author = {Li, Xiaochen and Wu, Peng and He, Yufeng and Chen, Kai},
Title = {A Compact Seafloor Electromagnetic Receiver for Ship Electromagnetic
   Signature Measurement},
Journal = {IEEE TRANSACTIONS ON INSTRUMENTATION AND MEASUREMENT},
Year = {2025},
Volume = {74},
DOI = {10.1109/TIM.2025.3572163},
Article-Number = {9523611},
ISSN = {0018-9456},
EISSN = {1557-9662},
Unique-ID = {WOS:001512566700021},
}

@article{ Yang2025,
Author = {Yang, Chenglong and Dai, Wenshuo and Zhang, Hengyou and Yu, Shengbao and
   Zhang, Xinhao},
Title = {Three-Component Coupled Coil Transmission System With Reactance
   Cancellation for MCSEM Prospecting},
Journal = {IEEE TRANSACTIONS ON INSTRUMENTATION AND MEASUREMENT},
Year = {2025},
Volume = {74},
DOI = {10.1109/TIM.2025.3533647},
Article-Number = {2001814},
ISSN = {0018-9456},
EISSN = {1557-9662},
ResearcherID-Numbers = {Zhang, Xinhao/KHY-1758-2024
   },
ORCID-Numbers = {Zhang, Hengyou/0009-0005-0013-7220
   Zhang, Xinhao/0000-0002-8280-0217
   Dai, Wen shuo/0009-0004-0465-5802},
Unique-ID = {WOS:001416246500031},
}

@article{ Kemp2019,
Author = {Kemp, Mark A. and Franzi, Matt and Haase, Andy and Jongewaard, Erik and
   Whittaker, Matthew T. and Kirkpatrick, Michael and Sparr, Robert},
Title = {A high Q piezoelectric resonator as a portable VLF transmitter},
Journal = {NATURE COMMUNICATIONS},
Year = {2019},
Volume = {10},
Month = {APR 12},
DOI = {10.1038/s41467-019-09680-2},
Article-Number = {1715},
ISSN = {2041-1723},
ResearcherID-Numbers = {Kirkpatrick, Michael/OYE-0128-2025
   Whittaker, Matt/KIA-6564-2024},
Unique-ID = {WOS:000464338100026},
}

@inproceedings{ Li2016,
Author = {Li Lihua and Wang Yongbin and Liu Shangfu and Liu Tong},
Book-Group-Author = {IEEE},
Title = {A Channel Coding Schematic Design of Super Low Frequency Communication},
Booktitle = {2016 FIRST IEEE INTERNATIONAL CONFERENCE ON COMPUTER COMMUNICATION AND
   THE INTERNET (ICCCI 2016)},
Year = {2016},
Pages = {66-69},
Note = {1st IEEE International Conference on Computer Communication and the
   Internet (ICCCI), Wuhan, PEOPLES R CHINA, OCT 13-15, 2016},
Organization = {IEEE},
ISBN = {978-1-4673-8515-2},
Unique-ID = {WOS:000390712100015},
}

@inproceedings{ Sacchi2003,
Author = {Sacchi, E and Bietti, I and Erba, S and Tee, L and Vilmercati, P and
   Castello, R},
Book-Group-Author = {IEEE
   IEEE},
Title = {A 15 mW, 70 kHz 1/f corner direct conversion CMOS receiver},
Booktitle = {PROCEEDINGS OF THE IEEE 2003 CUSTOM INTEGRATED CIRCUITS CONFERENCE},
Year = {2003},
Pages = {459-462},
Note = {25th Annual Custom Integrated Circuits Conference, SAN JOSE, CA, SEP
   21-24, 2003},
Organization = {SSCS; IEEE; EDS},
DOI = {10.1109/CICC.2003.1249440},
ISBN = {0-7803-7842-3},
ORCID-Numbers = {CASTELLO, RINALDO/0000-0002-8375-3862},
Unique-ID = {WOS:000186714300093},
}

@article{ SCOFIELD1994,
Author = {SCOFIELD, JH},
Title = {FREQUENCY-DOMAIN DESCRIPTION OF A LOCK-IN AMPLIFIER},
Journal = {AMERICAN JOURNAL OF PHYSICS},
Year = {1994},
Volume = {62},
Number = {2},
Pages = {129-133},
Month = {FEB},
DOI = {10.1119/1.17629},
ISSN = {0002-9505},
Unique-ID = {WOS:A1994MU54000010},
}

@article{ Sanduleanu2025,
Author = {Sanduleanu, Mihai},
Title = {Fundamentals of 1/f noise reduction technique based on complementary
   cascode switching applied to a 52 microwatts 5.2pJ/bit 100Kb/s 120 GHz
   receiver},
Journal = {SCIENTIFIC REPORTS},
Year = {2025},
Volume = {15},
Number = {1},
Month = {JUL 1},
DOI = {10.1038/s41598-025-08516-y},
Article-Number = {22291},
ISSN = {2045-2322},
Unique-ID = {WOS:001523016900030},
}

@article{ Enz1996,
Author = {Enz, CC and Temes, GC},
Title = {Circuit techniques for reducing the effects of op-amp imperfections:
   Autozeroing, correlated double sampling, and chopper stabilization},
Journal = {PROCEEDINGS OF THE IEEE},
Year = {1996},
Volume = {84},
Number = {11},
Pages = {1584-1614},
Month = {NOV},
DOI = {10.1109/5.542410},
ISSN = {0018-9219},
ResearcherID-Numbers = {Enz, Christian/PGA-1372-2026},
Unique-ID = {WOS:A1996VQ88900004},
}

@article{ ENZ1987,
Author = {ENZ, CC and VITTOZ, EA and KRUMMENACHER, F},
Title = {A CMOS CHOPPER AMPLIFIER},
Journal = {IEEE JOURNAL OF SOLID-STATE CIRCUITS},
Year = {1987},
Volume = {22},
Number = {3},
Pages = {335-342},
Month = {JUN},
DOI = {10.1109/JSSC.1987.1052730},
ISSN = {0018-9200},
ResearcherID-Numbers = {Enz, Christian/PGA-1372-2026},
Unique-ID = {WOS:A1987H276300005},
}

@article{Jing2020,
   author = {Jing, M. Y. and Hu, Y. and Ma, J. and Zhang, H. and Zhang, L. J. and Xiao, L. T. and Jia, S. T.},
   title = {Atomic superheterodyne receiver based on microwave-dressed Rydberg spectroscopy},
   journal = {Nature Physics},
   volume = {16},
   number = {9},
   pages = {911-+},
   ISSN = {1745-2473},
   DOI = {10.1038/s41567-020-0918-5},
   year = {2020},
   type = {Journal Article}
}

@article{Hu2022,
   author = {Hu, J. L. and Li, H. Q. and Song, R. and Bai, J. X. and Jiao, Y. C. and Zhao, J. M. and Jia, S. T.},
   title = {Continuously tunable radio frequency electrometry with Rydberg atoms},
   journal = {Applied Physics Letters},
   volume = {121},
   number = {1},
   pages = {6},
   ISSN = {1077-3118
0003-6951},
   DOI = {10.1063/5.0086357},
   year = {2022},
   type = {Journal Article}
}

@article{Meyer2020,
   author = {Meyer, D. H. and Castillo, Z. A. and Cox, K. C. and Kunz, P. D.},
   title = {Assessment of Rydberg atoms for wideband electric field sensing},
   journal = {Journal of Physics B-Atomic Molecular and Optical Physics},
   volume = {53},
   number = {3},
   pages = {12},
   ISSN = {0953-4075},
   DOI = {10.1088/1361-6455/ab6051},
   year = {2020},
   type = {Journal Article}
}

@article{Chu1948,
   author = {Chu, L. J.},
   title = {PHYSICAL LIMITATIONS OF OMNI-DIRECTIONAL ANTENNAS},
   journal = {Journal of Applied Physics},
   volume = {19},
   number = {12},
   pages = {1163-1175},
   ISSN = {0021-8979},
   DOI = {10.1063/1.1715038},
   year = {1948},
   type = {Journal Article}
}

@article{Jau2020,
   author = {Jau, Y. Y. and Carter, T.},
   title = {Vapor-Cell-Based Atomic Electrometry for Detection Frequencies below 1 kHz},
   journal = {Physical Review Applied},
   volume = {13},
   number = {5},
   pages = {11},
   ISSN = {2331-7019},
   DOI = {10.1103/PhysRevApplied.13.054034},
   year = {2020},
   type = {Journal Article}
}

@article{Abel2011,
   author = {Abel, R. P. and Carr, C. and Krohn, U. and Adams, C. S.},
   title = {Electrometry near a dielectric surface using Rydberg electromagnetically induced transparency},
   journal = {Physical Review A},
   volume = {84},
   number = {2},
   pages = {5},
   ISSN = {1050-2947},
   DOI = {10.1103/PhysRevA.84.023408},
   year = {2011},
   type = {Journal Article}
}

@article{Bohaichuk2022,
   author = {Bohaichuk, S. M. and Booth, D. and Nickerson, K. and Tai, H. R. Y. and Shaffer, J. P.},
   title = {Origins of Rydberg-Atom Electrometer Transient Response and Its Impact on Radio-Frequency Pulse Sensing},
   journal = {Physical Review Applied},
   volume = {18},
   number = {3},
   pages = {15},
   ISSN = {2331-7019},
   DOI = {10.1103/PhysRevApplied.18.034030},
   year = {2022},
   type = {Journal Article}
}

@article{Anderson2021,
   author = {Anderson, D. A. and Sapiro, R. E. and Raithel, G.},
   title = {A Self-Calibrated SI-Traceable Rydberg Atom-Based Radio Frequency Electric Field Probe and Measurement Instrument},
   journal = {Ieee Transactions on Antennas and Propagation},
   volume = {69},
   number = {9},
   pages = {5931-5941},
   ISSN = {0018-926X},
   DOI = {10.1109/tap.2021.3060540},
   year = {2021},
   type = {Journal Article}
}

@article{Gordon2019,
   author = {Gordon, J. A. and Simons, M. T. and Haddab, A. H. and Holloway, C. L.},
   title = {Weak electric-field detection with sub-1 Hz resolution at radio frequencies using a Rydberg atom-based mixer},
   journal = {Aip Advances},
   volume = {9},
   number = {4},
   pages = {5},
   DOI = {10.1063/1.5095633},
   year = {2019},
   type = {Journal Article}
}

@article{ Yan2025,
Author = {Yan, Shaochen and Li, Xinyao and Pang, Xiaoyan and Wang, Ruiqiong and
   Wen, Wen and Zhai, Weile and Cui, Wanzhao and Gao, Yongsheng},
Title = {Measurement of Doppler Frequency Shift and Angle of Arrival With Rydberg
   Atom-Based Sensors},
Journal = {IEEE TRANSACTIONS ON INSTRUMENTATION AND MEASUREMENT},
Year = {2025},
Volume = {74},
DOI = {10.1109/TIM.2025.3568981},
Article-Number = {1505409},
ISSN = {0018-9456},
EISSN = {1557-9662},
ResearcherID-Numbers = {Pang, Xiaoyan/M-9413-2013
   Gao, Yongsheng/I-5174-2017
   },
ORCID-Numbers = {Zhai, Weile/0000-0001-6015-9804
   WANG, RUIQIONG/0009-0009-9565-9433
   Pang, Xiaoyan/0000-0001-8265-0952
   Yan, Shaochen/0009-0007-9903-6175
   Gao, Yongsheng/0000-0002-0388-0096
   wanzhao, cui/0000-0002-1636-2925},
Unique-ID = {WOS:001497993500012},
}

@article{Li2023,
   author = {Li, L. and Jiao, Y. C. and Hu, J. L. and Li, H. Q. and Shi, M. and Zhao, J. M. and Jia, S. T.},
   title = {Super low-frequency electric field measurement based on Rydberg atoms},
   journal = {Optics Express},
   volume = {31},
   number = {18},
   pages = {29228-29234},
   ISSN = {1094-4087},
   DOI = {10.1364/oe.499244},
   year = {2023},
   type = {Journal Article}
}

@article{Lei2024,
   author = {Lei, M. W. and Shi, M.},
   title = {High sensitivity measurement of ULF, VLF, and LF fields with a Rydberg-atom sensor},
   journal = {Optics Letters},
   volume = {49},
   number = {19},
   pages = {5547-5550},
   ISSN = {0146-9592},
   DOI = {10.1364/ol.539090},
   year = {2024},
   type = {Journal Article}
}

@article{Zhang2026,
   author = {Zhang, Jinhao and Sun, Zhanshan and Yao, Jiawei and Zhao, Fengting and Lin, Yi and Sang, Di and Yang, Kai and An, Qiang and Fu, Yunqi},
   title = {Self-dressing Rydberg atomic receiver based on laser-induced DC field},
   journal = {npj Quantum Materials},
   ISSN = {2397-4648},
   DOI = {10.1038/s41535-026-00862-y},
   year = {2026},
   type = {Journal Article}
}

@article{Mao2024,
   author = {Mao, Ruiqi and Lin, Yi and Zhou, Aojie and Yang, Kai and Fu, Yunqi %J IEEE Transactions on Antennas and Propagation},
   title = {Shortwave Ultrahigh-Sensitivity Rydberg Atomic Electric Field Sensing Based on a Subminiature Resonator},
   number = {11},
   pages = {72},
   year = {2024},
   type = {Journal Article}
}

@article{Jiao2017,
   author = {Jiao, Y. C. and Hao, L. P. and Han, X. X. and Bai, S. Y. and Raithel, G. and Zhao, J. M. and Jia, S. T.},
   title = {Atom-Based Radio-Frequency Field Calibration and Polarization Measurement Using Cesium <i>nD<sub>J</sub></i> Floquet States},
   journal = {Physical Review Applied},
   volume = {8},
   number = {1},
   pages = {7},
   ISSN = {2331-7019},
   DOI = {10.1103/PhysRevApplied.8.014028},
   year = {2017},
   type = {Journal Article}
}

@article{Sibalic2017,
   author = {Sibalic, N. and Pritchard, J. D. and Adams, C. S. and Weatherill, K. J.},
   title = {ARC: An open-source library for calculating properties of alkali Rydberg atoms},
   journal = {Computer Physics Communications},
   volume = {220},
   pages = {319-331},
   ISSN = {0010-4655},
   DOI = {10.1016/j.cpc.2017.06.015},
   year = {2017},
   type = {Journal Article}
}

@article{ Arumugam2025,
Author = {Arumugam, Darmindra},
Title = {Stark modulated Rydberg dissipative time crystals at room temperature
   applied to sub-kHz electric field sensing},
Journal = {SCIENTIFIC REPORTS},
Year = {2025},
Volume = {15},
Number = {1},
Month = {OCT 15},
DOI = {10.1038/s41598-025-19859-x},
Article-Number = {35976},
ISSN = {2045-2322},
Unique-ID = {WOS:001596279600013},
}

@article{Hansen2006,
  title={Electrically Small, Superdirective, and Superconducting Antennas},
  author={ Hansen, R. },
  year={2006},
}





\end{document}